\documentclass[prc,amsmath,twocolumn,showkeys,showpacs,superscriptaddress]{revtex4}

\usepackage{latexsym}
\usepackage{amssymb}
\usepackage{dsfont}
\usepackage{amsbsy}
\usepackage{amsmath}
\usepackage[varg]{txfonts}
\usepackage{mathrsfs}
\usepackage{upgreek}
\usepackage{verbatim}
\usepackage{hyperref}
\newsavebox{\fmbox}

\newcommand{\braket}[1]{\left\langle {#1} \right\rangle }
\newcommand{\ket}[1]{\left|{#1} \right\rangle }
\newcommand{\bra}[1]{\left\langle {#1}\right|}
\newcommand\idop{\mathds 1}
\DeclareMathAlphabet{\mathpzc}{OT1}{pzc}{m}{it}

\usepackage{graphicx,color}
\usepackage{amssymb}
\usepackage{enumerate}
\usepackage{verbatim}
\usepackage{natbib}
\usepackage{color}
\usepackage{array}

\begin{document}
  \newcommand {\nc} {\newcommand}
  \nc {\Sec} [1] {Sec.~\ref{#1}}
  \nc {\IR} [1] {\textcolor{red}{#1}} 
  \nc {\IB} [1] {\textcolor{blue}{#1}}

\title{Establishing a theory for deuteron induced surrogate reactions}

\author{G. Potel}
\affiliation{National Superconducting Cyclotron Laboratory, Michigan State University, East Lansing, Michigan 48824, USA}
\affiliation{Lawrence Livermore National Laboratory L-414, Livermore, CA 94551, USA}
\author{F.~M.~Nunes}
\affiliation{National Superconducting Cyclotron Laboratory, Michigan State University, East Lansing, Michigan 48824, USA}
\affiliation{Department of Physics and Astronomy, Michigan State University, East Lansing, MI 48824-1321}
\author{I.J. Thompson}
\affiliation{Lawrence Livermore National Laboratory L-414, Livermore, CA 94551, USA}

\date{\today}

%%%%%%%%%%%%%%%%%%%%%%%%%%%%%%%%%%%%%%%%%%%%%%%%%%%%%%%%%%%%%%%%%%%%%%%%%%%%%%%%%%%%%%%%%%%%%%%%%%%%%%%%%%%%%%%%%%%%%%%%%%%%%%%%%%%

\begin{abstract}
{\bf Background:} Deuteron induced reactions serve as surrogates for neutron capture into compound states. Although these reactions are of great applicability, no theoretical efforts have been invested in this direction over the last decade.
{\bf Purpose:} The goal of this work is to establish on firm grounds a theory for deuteron induced neutron capture reactions. This includes formulating elastic and inelastic breakup in a consistent manner.
{\bf Method:}  We describe this process both in post and prior form distorted wave Born approximation following previous works and discuss the differences in the formulation. While the convergence issues arising in the post formulation can be overcome in the prior formulation, in this case one still needs to take into account additional terms due to non-orthogonality.
{\bf Results:}  We apply our method to the $^{53}$Nb(d,p)X at $E_d=$15 MeV and 25 MeV, and are able to obtain a good description of the data. We look at the various partial wave contributions, as well as elastic versus inelastic contributions. We also connect our formulation with transfer to neutron bound states. 
{\bf Conclusions:}  Our calculations demonstrate that the non-orthogonality term arising in the prior formulation is significant and is at the heart of the long-standing controversy between the post and the prior formulations of the theory. We also show that the cross sections for these reactions are angular momentum dependent and therefore the commonly used Weisskopf limit is inadequate. Finally we make important predictions for the relative contributions of elastic breakup and non-elastic breakup, and call for elastic breakup measurements to further constrain our model.
\end{abstract}



\maketitle
\section{Introduction}
\label{intro}

Neutron capture reactions A(n,$\gamma$)B are very important in astrophysics, for the production of  heavy elements, but are equally relevant in stewardship science, since it is through neutron capture that fission is induced and chain reactions begin. Most often, for low energy neutrons, the capture proceeds through continuum states, forming a compound nucleus (A+n$\rightarrow$ B$^*$) that then decays, either through gamma emission to the ground state, or through other particle channels. The direct measurement of neutron capture is challenging, particularly because most of the targets of interest have short half-lives and neutrons cannot be made into targets. A proposed alternative it to use deuterons as surrogates \cite{escher2012}. In this indirect method, the proton inside the deuteron behaves mostly as a spectator, the neutron inside the deuteron is delivered to the target surface and gets absorbed by the target A(d,p)B$^*$. Since this is a compound nucleus reaction, the final compound nucleus decays in the same way as after A+n$\rightarrow$B$^*$. An example of a recent measurement applying the surrogate method is the $^{171,173}$Yb(d,p$\gamma$) experiment performed at LBNL \cite{hatarik2010}.  The measurement was performed on a nucleus for which neutron capture cross sections were already available. The neutron capture cross-sections extracted in the deuteron induced reactions were in fair agreement with those measured directly. 

While the exclusive process A(d,pn)A (with A left in its ground state) is clearly identified as elastic breakup, the rest of the cross section arising from the inclusive process A(d,p)X, where only the proton is measured in the final state, is harder to name because it encompasses many different processes. Some refer to this component of the cross section as inelastic breakup, breakup-fusion or partial fusion. In this work, we will always use the term non-elastic breakup.

Although there are important applications of the surrogate method for neutron capture, no theoretical development has taken place in the last two decades to establish the method on firm grounds. In terms of direct reaction mechanisms, this process can be seen as inelastic breakup followed by fusion, or neutron transfer to the continuum. Significant theoretical effort took place in the eighties with the main idea being first introduced by Kerman and McVoy \cite{kerman1979}, with the works of  Udagawa and Tamura \cite{udagawa1980} and Austern and Vincent \cite{austern1981} appearing shortly after. In \cite{udagawa1980} the authors assume the process A(d,p)B$^*$ is a two step process, first breakup of the deuteron followed by fusion of the neutron. They describe this in distorted wave Born approximation in prior form and make a number of additional approximations (we will denote this theory as UT). In \cite{austern1981},  the starting point is the post-form distorted wave Born approximation, and the authors assume that the target gets excited only by the neutron-target interaction (we will denote this theory as AV). One difficulty in this method is the convergence of the matrix element. In both \cite{austern1981} and \cite{udagawa1980},  all (d,p) transfer cross sections to the excited states are summed without explicitly introducing the properties of these states. This is of course a key aspect in describing inclusive processes. As it turned out, cross sections obtained with the post (AV) theory did not agree with those obtained with the prior (UT) theory. This generated a heated controversy that lasted a decade and was never fully resolved. 

A detailed analysis of the UT theory is presented in \cite{li1984}, where several approximations are considered including the zero range approximation and the surface approximation. The authors of \cite{li1984,udagawa1986}  argue that the theory of AV includes unphysical components which should be corrected by inclusion of a non-orthogonality term. On the other hand, Ichimura et al. \cite{ichimura1985} in their detailed examination, conclude that certain implicit approximations made in the optical reductions by UT are at the heart of the disagreement. Although several groups revisited the matter later \cite{ichimura1990,Hussein:85,hussein89,hussein90}, establishing a relationship between the various theories, the controversy on the relevance of the non-orthogonality term was never resolved. Nowadays, many of the approximations made in the early eighties have become unnecessary, and thus it makes sense to revisit the issue. 

Two recent works have applied the AV theory to study a variety of reactions \cite{lei2015,carlson2015}. In Lei and Moro \cite{lei2015}  these reactions include deuteron breakup on $^{58}$Ni at intermediate energies, deuteron breakup on $^{93}$Nb at lower energies and $^6$Li elastic scattering on $^{209}$Bi around the Coulomb barrier. In order to deal with the convergence of the amplitude, the authors in \cite{lei2015} construct continuum bins corresponding to a square integrable wave packet obtained averaging the scattering states over energy.  These studies \cite{carlson2015,lei2015} show that, overall, the AV theory provides a good description of the processes considered.

Given the recent experimental interest in the surrogate method, it is critical to develop a theory that is practical and reliable.
Our overall goal with this work is exactly to establish such a theory, which can then serve as a starting point to make further improvements in the future. In this work we will present our own derivations of the elastic and inelastic breakup amplitudes in both the post (AV) and the prior (UT) forms, within the distorted wave Born approximation (section II). We will show that these two theories are indeed equivalent if no further approximations are made, although the post formalism introduces numerical difficulties which are avoidable by the prior formalism. We will also show that a non-orthogonality term and additional cross-terms naturally arise in the UT theory.  This is a consequence of the fact that when writing the amplitude in the prior form,  no easy separation between the breakup process and the excitation process is possible.  In other words, in the prior formalism, breakup and excitation are entangled. These days, however, their computation poses no difficulty.  Following the work of \cite{udagawa1987}, in section III, we will use the same framework to study neutron transfer to bound states. This establishes an important connection between scattering and bound states, and provides a stringent test for the theory. We apply the method to surrogate reactions of deuterons on $^{93}$Nb for beam energies in the range of 10-20 MeV to compare with data, as done by \cite{mastroleo1990} (section V). We analyze our results and  dissect the various contributions, in terms of angular momentum, as well as the relative magnitude of elastic versus non-elastic cross sections. Finally, in Section VI, we draw our conclusions and provide an outlook into further possible developments.

%%%%%%%%%%%%%%%%%%%%%%%%%%%%%%%%%%%%%%%%%%%%%%%%%%%%%%%%%%%%%%%%%%%%%%%%%%%%%%%%%%%%%%%%%%%%%%%%%
\section{Theoretical formulation}
\label{theory}

\subsection{General formalism in the \emph{post} representation (AV)}
\label{theory-AV}

Let us consider the reaction A(d,p)B* which includes elastic breakup and any other inelastic processes. In this section, we follow closely the work by Austern {\it et al.} \cite{austern1981} and derive the post-form amplitude for the process. We will adopt a spectator approximation for the proton, which means the proton--target interaction will not explicitly excite the target $A$. We will thus start by substituting the proton--target interaction $V_{Ap}(r_{Ap},\xi_A)$ by an optical potential $U_{Ap}(r_{Ap})$. In addition, for the purpose of our derivation, we have considered $A$ to be infinitely massive. However, in the actual numerical applications, the recoil of the nucleus $A$ is fully taken into account.

The three-body Hamiltonian for the problem is
\begin{align}\label{eq24}
H=K_n+K_p+h_A(\xi_A)+V_{pn}(r_{pn})+V_{An}(r_{An},\xi_A)+U_{Ap}(r_{Ap}),
\end{align}
where $K_n$ and $K_p$ are the kinetic energy operators acting on the neutron and  proton coordinates respectively.  We now consider the model wavefunction
\begin{align}\label{eq135}
\Psi=\chi_i\phi_d\phi_A+\chi_f\phi_B^c,
\end{align}
where $\phi_d$ is the deuteron eigenfunction, $\phi_A$ is the ground state of the target nucleus, and $\phi_B^c$ represents the $c$'th eigenstate of the final compound nucleus $B$. These wave functions satisfy
\begin{align}\label{eq138}
\nonumber & h_d\phi_d=\left(K_{np}+V_{np}\right)\,\phi_d=\varepsilon_d\phi_d,\\
& h_A\phi_A=\varepsilon_A\phi_A.\\
\nonumber & h_B\phi_B^c=\left(K_n+h_A+V_{An}\right)\,\phi_B^c=\varepsilon_B^c\phi_B^c.
\end{align}
Here, $K_{np}$ is the  kinetic energy of the neutron--proton motion, and $h_A$ is the internal Hamiltonian of the target nucleus A.
\begin{figure}[t!]
\begin{center}
\includegraphics[width=0.6\columnwidth]{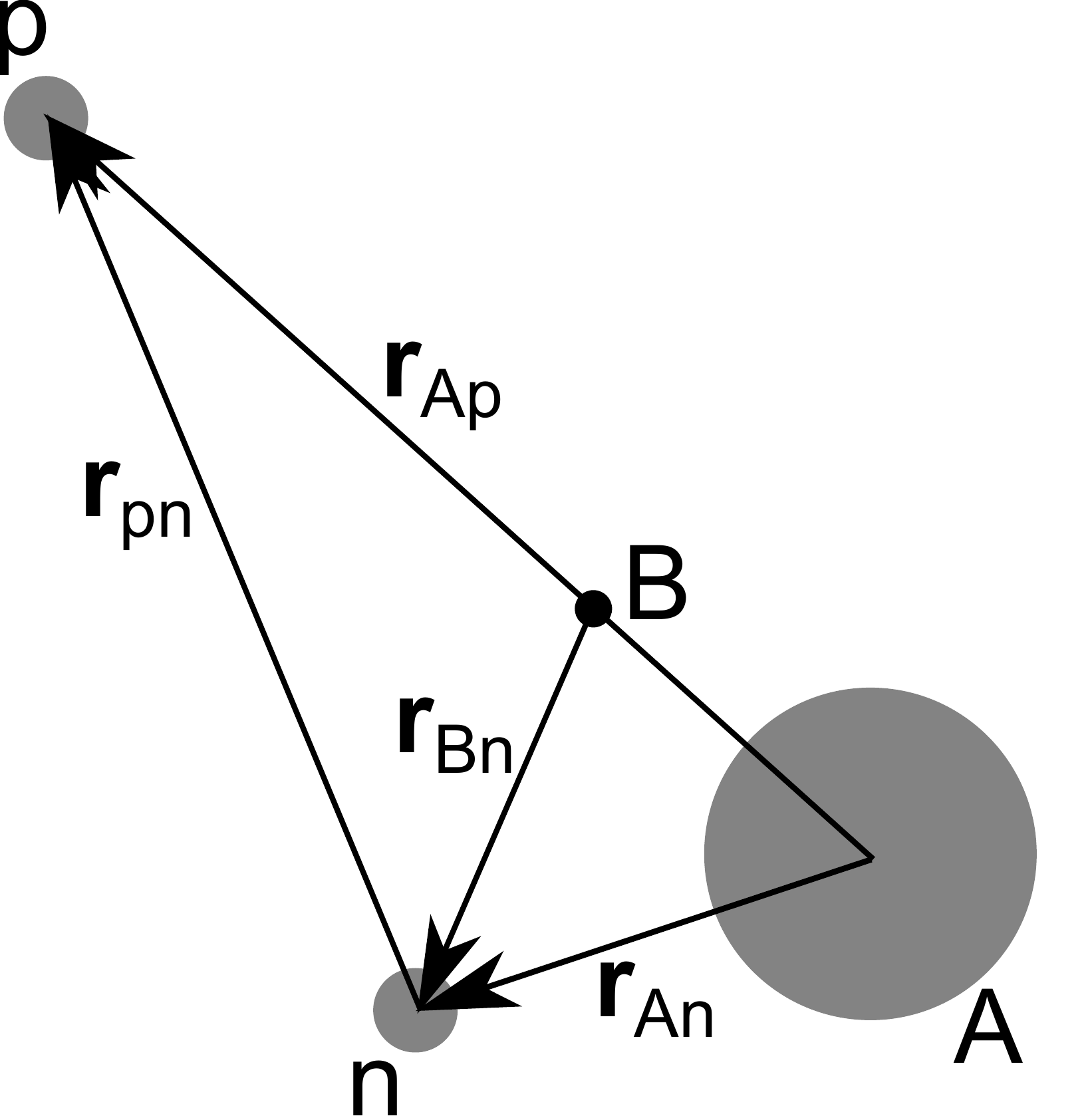}
\end{center}
\caption{Definition of the coordinates used in our formulations.}
\label{fig:coord}
\end{figure} 
In Fig.\ \ref{fig:coord} we define the coordinates that we use in our formulation. These will become useful throughout this section.

Considering only first order in the couplings between the incoming deuteron channel $i$ and the final proton channel $f$, the coupled equations for the unknowns $\chi_i$ and $\chi_f$ simplify to the  distorted wave Born approximation (DWBA) differential equations, and can be written in either prior and post form \cite{book},
\begin{align}
\label{eq136}
\left(E_i-H_i\right)\chi_i&=0\\
\nonumber \left(E_f-H_f\right)\chi_f&=\left\{\begin{array}{l}
 \braket{\phi_B^c|V_{prior}|\phi_d\phi_A}\chi_i\\ 
\braket{\phi_B^c|V_{post}|\phi_d\phi_A}\chi_i+\left(H_f-E_f\right)\braket{\phi_B^c|\phi_d\phi_A}\chi_i,
\end{array} \right. 
\end{align}
where the initial and final Hamiltonians are 
\begin{align}
\label{eq-hihf}
\nonumber & H_i=K_d+U_{Ad}   \\
& H_f=K_p+U_{Ap}, 
\end{align}
with the prior and post operators 
\begin{align}
\label{eq137}
\nonumber & V_{prior}=V_{An}+U_{Ap}-U_{Ad}, \\
& V_{post}=V_{np} .
\end{align}
and the energies $E_i=E-\varepsilon_d-\varepsilon_A$ and $E_f=E-\varepsilon_B^c$. For a target $A$ with a big but finite mass,  the \textit{post} form will include a negligible remnant term $U_{Ap}-U_{Bp}$, $U_{Bp}$ being the optical potential between the proton and the nucleus $B$. In the approximation in which the proton is treated as a spectator, it is important to note that, in any case, this term does not depend upon the intrinsic coordinates of the target $A$.

According to eq. (\ref{eq136}), an auxiliary wavefunction $X_f$  defined by
 \begin{align}\label{eq146} 
X_f=\chi_f+\braket{\phi_B^c|\phi_d\phi_A}\chi_i,
 \end{align}
is such that, in the post representation,
  \begin{align}\label{eq147}
 \left(H_f-E_f\right) X_f=\braket{\phi_B^c|V_{post}|\phi_d\phi_A}\chi_i.
  \end{align}
The non--orthogonality term $\braket{\phi_B^c|\phi_d\phi_A}$ vanishes for large values of the proton coordinate, so $X_f$ and $\chi_f$ are asymptotically identical. Therefore, the $T$-matrix $T_{dc}$ for the detection of the proton after the population of the state $\phi_B^c$ of the residual nucleus has two equivalent expressions:
\begin{align}\label{eq139}
T_{dc}=\braket{\chi_f^{(-)}\phi_B^c|V_{prior}|\phi_d\phi_A\chi_i}=\braket{\chi_f^{(-)}\phi_B^c|V_{post}|\phi_d\phi_A\chi_i},
\end{align}
and the celebrated \textit{post}--\textit{prior} equivalence in  DWBA  is verified. The proton distorted wave $\chi_f^{(-)}$ satisfies the equation:
\begin{align}\label{eq140}
\left(E_f-H^\dagger_f\right)\chi_f^{(-)}=0.
\end{align}

The inclusive cross section is summed over all $B$ channels of a given energy. The inclusive breakup cross section for the process $d+A\rightarrow p+B$ is thus 
\begin{align}\label{eq25}
\nonumber &\frac{d^2\sigma}{d\Omega_pdE_p}=\frac{2\pi}{\hbar v_d}\rho_p(E_p)\sum_c\left|\left\langle\chi_f^{(-)}\phi_B^c|V\,|\,\phi_d\phi_A\chi_i\right\rangle\right|^2\delta(E{-}E_p{-}\varepsilon_B^c)\\
~~&=-\frac{2}{\hbar v_d}\rho_p(E_p)\Im\left\langle\,\phi_d\phi_A\chi_i|V^\dagger\,|\chi_f^{(-)}\right) G_B\left(\chi_f^{(-)}|V\,|\,\phi_d\phi_A\chi_i\right\rangle,
\end{align}
 where the round brackets indicate that we only integrate over the proton coordinate. Here,  $V$ stands for either $V_{prior}$ or $V_{post}$, 
  \begin{align}\label{eq82}
 \rho_p(E_p)=\frac{m_p k_p}{8\pi^3\hbar^2}
  \end{align}
  is the proton level density, and $E_p$ is the kinetic energy of the detected proton.
  
The energy--conserving delta function $\delta(E-E_p-\varepsilon_B^c)$  can be written as the imaginary part of the spectral expansion
\begin{align}\label{eq26}
G_{B}=\lim_{\epsilon\rightarrow 0}\sum_c\frac{\ket{\phi_B^c}\bra{\phi_B^c}}{E-E_p-\varepsilon_B^c+i\epsilon}.
\end{align}
This Green's function, in operator form,  is simply
\begin{align}\label{eq26B}
G_B = \frac{1}{ E - E_p - h_B+i\epsilon}.
\end{align}

Since $V_{np}$ is not contained in $h_B$, Eq.(\ref{eq26B}) can be interpreted as a propagator in the breakup channel, i.e., when the deuteron bound state is  absent. 
Furthermore, if we choose the {\em post} representation of Eq.(\ref{eq25}), we have the advantage that the potential $V=V_{post}=V_{pn}(r_{pn})$  does not depend on $\xi_A$, so we can directly project the Green's function onto the target ground  state, the so-called optical reduction,
\begin{align}\label{eq31}
G_B^{opt}=\langle \phi_A|G_B|\phi_A\rangle=\frac{1}{E-E_p{-}\varepsilon_A-K_n-U_{An}(r_{An})+i\epsilon}.
\end{align}
The exact form of $U_{An}$ obtained from the above projection operation is the Feshbach optical potential for the $n-A$ system in the ground state of $A$, and includes  the effects of all possible excitations of $A$ due to the interaction with the neutron. In the present paper we use, instead, a global optical potential obtained from a systematic fitting to elastic scattering data \cite{koning03}. For both positive and negative energies, this optical potential $U_{An}=V_{An}+iW_{An}$ (with $V_{An},W_{An}$ real) should be interpreted as an energy--averaged approximation to the exact Feshbach potential.
%Here   $U_{An}=V_{An}+iW_{An}$ (with $V_{An},W_{An}$ real) is the Feshbach optical potential for the $n-A$ system in the ground state of $A$, and includes implicitly the effects of all possible excitations of $A$ due to the interaction with the neutron.

The inclusive cross section has thus the post-form expression
\begin{align}\label{eq32}
\frac{d^2\sigma}{d\Omega_pdE_p}=-\frac{2}{\hbar v_d}\rho_p(E_p)\Im\left\langle\vphantom{\chi_f}\,\phi_d\chi_i\right|V_{pn}\,\left|\chi_f^{(-)}\right) G_B^{opt}\left(\chi_f^{(-)}\right|V_{pn}\,\left|\,\vphantom{\chi_f}\phi_d\chi_i\right\rangle.
\end{align} 
Because the real potential $V_{np}(r_{np})$  cannot excite the nucleus $A$, this formalism suggests a two--step mechanism: $V_{np}(r_{np})$ first breaks up the deuteron and then $G_B^{opt}$ propagates the system in the breakup channel, eventually leading to the absorption of the neutron in the complex field $U_{An}$. 

It is useful to extract from Eq.(\ref{eq32}) the contributions of breakup without the excitation of the target $A$ (elastic breakup, EB), and  non-elastic breakup (NEB) where the target no longer remains in its ground state. To this purpose, we transform
\begin{align}\label{eq148}
\nonumber G_B^{opt}=&G_0\left(1+G_B^{opt}U_{An}\right)\\
\nonumber &=(1+G_B^{opt\dagger} U_{An}^\dagger)\, G_0\left(1+G_B^{opt}U_{An}\right)-G_B^{opt\dagger} U_{An}^\dagger G_B^{opt},\\
G_B^{opt\dagger}
&=(1+G_B^{opt\dagger} U_{An}^\dagger)\, G_0^\dagger\left(1+G_B^{opt}U_{An}\right)-G_B^{opt^\dagger} U_{An} G_B^{opt},
\end{align}
where $G_0$ is the propagator in free space. 
 We now define the scattering waves $|\chi_n(r_n;k_n)\rangle$ subject to the optical potential $U_{An}$:
\begin{align}\label{eq149}
|\chi_n(r_{An};k_n)\rangle=(1+G_B^{opt} U_{An})\ket{\chi_0(r_{An};k_n)},
\end{align}
where $|\chi_0(r_{An};k_n)\rangle$ is a free  plane wave with momentum $k_n$.
From these expressions, we obtain
\begin{align}\label{eq33}
\nonumber\Im G_B^{opt}=&(1+G_B^{opt^\dagger} U_{An}^\dagger)\,\Im G_0\left(1+G_B^{opt}U_{An}\right)+G_B^{opt^\dagger} W_{An} G_B^{opt}\\
\nonumber =&-\pi\sum_{k_n}|\chi_n(r_n;k_{An})\rangle\delta\left(E-E_p-\frac{k_n^2}{2m_n}\right)\langle\chi_n(r_{An};k_n)|  \\
&+G_B^{opt\dagger} W_{An}\, G_B^{opt}.
\end{align}

The first term of Eq.(\ref{eq148}), when used in Eq.(\ref{eq32}), gives rise to  DWBA  elastic breakup.
For this we have  transition matrix elements
  \begin{align}\label{eq126} T_{EB}(\mathbf{k}_n,\mathbf{k}_p)=\left\langle\chi_f^{(-)}\chi_n\right|V_{pn}\,\left|\,\vphantom{\chi_f}\phi_d\,\chi_i\right\rangle.
  \end{align}
  The elastic-breakup differential cross section for the protons is given by integrating over the neutron angles:
  \begin{align}\label{eq133}
  \left.\frac{d^2\sigma}{dE_pd\Omega_{p}}\right]^{EB}=\frac{2\pi}{\hbar v_d}\rho_p(E_p)\rho_n(E_n)
     \int  |T_{EB}(\mathbf{k}_n,\mathbf{k}_p)   |^2\,d\Omega_{Bn}, 
  \end{align}
where,  in conjunction with the proton density introduced in Eq.(\ref{eq82}), we  use the neutron density of states
  \begin{align}\label{eq134}
 \rho_n(E_n)=\frac{m_n k_n}{8\pi^3\hbar^2}.
  \end{align}

  The second term of Eq.(\ref{eq33}) gives rise to  the non-elastic breakup cross section. Defining the post form of the source term,
      \begin{align}\label{eq141}
 S_{post}=\left(\chi_f^{(-)}\right|V_{pn}\,\left|\,\vphantom{\chi_f}\phi_d\chi_i\right\rangle,
      \end{align}
      and the neutron wavefunction
\begin{align}\label{eq35}
\left|\psi^{post}_{n}\right\rangle=G_B^{opt}S_{post},
\end{align}
the non--elastic breakup differential cross section can then be written as
 \begin{align}\label{eq34}
 \left.\frac{d^2\sigma}{d\Omega_pdE_p}\right]^{NEB}=-\frac{2}{\hbar v_d}\rho_p(E_p)\left\langle\vphantom{\chi_f}\psi^{post}_{n}\right|  W_{An}\, \left|\vphantom{\chi_f}\psi^{post}_{n}\right\rangle.
 \end{align}

Eq. (\ref{eq33})  suggests that the elastic and non--elastic breakup have a common origin. In Appendix A, we show that  flux conservation laws give  rise to separate elastic and non-elastic  terms in a consistent manner.

%%%%%%%%%%%%%%%%%%%%%%%%%%%%%%%%%%%%%%%%%%%%%%%%%%%%%%%%%%%%%%%%%%%%%%%%%%%%%%%%%%%%%%%%%%%%%%%%%

\subsection{General formalism in the \emph{prior} representation (UT)}
\label{theory-UT}
 
Computing the neutron wavefunction Eq.(\ref{eq35}) presents well known numerical difficulties. As the deuteron wavefunction $\phi_d$ and the \textit{post} potential $V_{pn}$ have the same argument, the   source term (\ref{eq141}) oscillates indefinitely as a function of the neutron coordinate,  and the integral in Eq.(\ref{eq34}) converges exceedingly slowly. There are methods to deal with this numerical problem. Alternatively we follow Udagawa and Tamura (UT) and revert to the {\em prior} representation. Let us first write the inclusive cross section in the prior representation,
\begin{align}\label{eq59}
 \frac{d^2\sigma}{d\Omega_pdE_p}& =-\frac{2}{\hbar v_d}\rho_p(E_p)  \nonumber \\
& \times \Im\left\langle\vphantom{\chi_p}\,\phi_d\phi_A\chi_i\right|V^\dagger_{prior}\,\left|\chi_f^{(-)}\right) G_B\left(\chi_f^{(-)}\right|V_{prior}\,\left|\,\vphantom{\chi_f}\phi_d\phi_A\chi_i\right\rangle,
\end{align}
where $V_{prior}=V_{An}+U_{Ap}-U_d$ is the prior interaction potential (see Eq.(\ref{eq137})).
 Now, however, the optical reduction defined by Eq. (\ref{eq31}) cannot be readily applied, because $V_{prior}$ acts on the intrinsic coordinates $\xi_A$ of the nucleus $A$ and  does not commute with $\phi_A$.

 We will thus split  $V_{prior}$ in a term that can induce breakup (and commutes with $\phi_A$) and a term that can excite $\phi_A$ (and commutes with $\chi_f^{(-)}$),
   \begin{align}\label{eq50}
\nonumber V_{prior}=&\left(G_B^{-1}+V_{prior}\right)-G_B^{-1}\\
&=\left(U_{Ap}-U_{Ad}-h_A-K_n-E_p+E\right)-G_B^{-1}.
   \end{align} 
Note that, when doing the averaging over the states of the target, some terms drop out. Introducing $\Delta U= U_{Ap} - U_{Ad}$, we can write:
   \begin{align}\label{eq53}
\nonumber    \left\langle\,\phi_A  |V_{prior}^\dagger\,|\chi_f\right) G_B\left(\chi_f^{(-)}|V_{prior}\,|\,\phi_A\right\rangle  & \\
\nonumber =\left(\Delta U^\dagger -K_n+E_n\right)\left|\chi_f^{(-)}\right) G_B^{opt} & \left(\chi_f^{(-)}\right|\left(\Delta U -K_n+E_n\right)  \\
    +\ |\chi_f^{(-)})(K_n-\bar V_{An}-E_n)(\chi_f^{(-)}|\ +\ & (\Delta U^\dagger + \Delta U ) |\chi_f^{(-)})(\chi_f^{(-)}| 
   \end{align}   
where $E_n=E-E_p-\varepsilon_A$ and $\bar V_{An}=\langle\phi_A|V_{An}|\phi_A\rangle$. When taking the imaginary part in Eq.(\ref{eq59}), the last  term in Eq.(\ref{eq53}) gives no contribution since it is a real operator. If we substitute $E_n-K_n$ with $U_{An}$, we have
   \begin{align}\label{eq54}
   \nonumber   \Im  \left\langle\,\phi_A|V_{prior}^\dagger\,|\chi_f^{(-)}\right)& G_B\left(\chi_f^{(-)}|V_{prior}\,|\,\phi_A\right\rangle = \\
   \nonumber    =&\left(\Delta U^\dagger +U_{An}\right)|\chi_f^{(-)}) G_B^{opt}(\chi_f^{(-)}|\left(\Delta U +U_{An}\right)\\
   &   +|\chi_f^{(-)})(U_{An}-\bar V_{An})(\chi_f^{(-)}|.
      \end{align}  
It is convenient to rewrite Eq.\ref{eq54} in the following form:
\begin{align}\label{eq55}
  \nonumber  \Im  \left\langle\,\phi_A|V_{prior}^\dagger\,|\chi_f^{(-)}\right)& G_B\left(\chi_f^{(-)}|V_{prior}\,|\,\phi_A\right\rangle\\
   \nonumber    =&\left(\Delta U +U_{An}\right)^\dagger|\chi_f^{(-)}) G_B^{opt}(\chi_f^{(-)}|\left(\Delta U +U_{An}\right)\\
\nonumber     +& 2iW_{An}|\chi_f^{(-)}) G_B^{opt}(\chi_f^{(-)}|\left(\Delta U +U_{An}\right) \\
+&|\chi_f^{(-)})(U_{An}-\bar V_{An})(\chi_f^{(-)}|.
      \end{align}  
The substitution of $E_n-K_n$ with $U_{An}$ might seem dubious, because, even if it is clear from the first line of Eq. (\ref{eq25}) that we are dealing with on-shell quantities, the second line seems to formally include off-shell terms. But, if we consider
$ E_n-K_n=G_B^{opt^{-1}}+U_{An}$,
we can see that the off-shell term $G_B^{opt^{-1}}$ gives a real contribution to Eq.(\ref{eq53}), and vanishes when taking the imaginary part in Eq.(\ref{eq59}).

It is now convenient to define the prior-form source term
   \begin{align}\label{eq57}
S_{\rm prior}=\left(\,\chi_f^{(-)} \big | \Delta U+U_{An} \big |\phi_d\,\chi_i\right\rangle.
   \end{align}
Note that the operator $\Delta U+U_{An}$ differs from the prior interaction defined in Eq.(\ref{eq137}), namely  the complex optical potential $U_{An}$ is used instead of the real interaction $V_{An}$ that causes core excitation. The prior-form neutron wave function is
\begin{align}\label{eq143}
\psi_{n}^{prior}  =G_B^{opt} S_{\rm prior}.
      \end{align}
We also need to introduce the non-orthogonality function:
   \begin{align}\label{eq45}
   \psi_n^{HM}=\left(\chi_f^{(-)}\right|\,\left.\,\vphantom{\chi_f}\phi_d\,\chi_i\right\rangle.
   \end{align}
This last expression defines the neutron ``source'' function  obtained by Hussein and McVoy (HM) using somewhat different approximations \cite{Hussein:85}. Now, inserting (\ref{eq55}) in (\ref{eq59}), we can arrive at:
   \begin{align}\label{eq56}
  \nonumber & \frac{d^2\sigma}{d\Omega_pdE_p}  =-\frac{2}{\hbar v_d}\varphi(E_p)\left[\Im\left\langle\,S_{\rm prior}|G_B^{opt}\,|S_{\rm prior}\right\rangle \right.\\ 
  & +2\Re\left\langle\,\psi_n^{{HM}}|W_{An}G_B^{opt}\,|S_{\rm prior}\right\rangle+\left.\left\langle\,\psi_n^{{HM}}|W_{An}\,|\psi_n^{{HM}}\right\rangle\right],
   \end{align}

We can apply the identity (\ref{eq33}) to remove from the first term of (\ref{eq56}), the elastic breakup contribution. The remaining terms represent the total non-elastic breakup cross section:
   \begin{align}\label{eq58}
  \nonumber & \left.\frac{d^2\sigma}{d\Omega_pdE_p}\right]^{NEB}=-\frac{2}{\hbar v_d}\rho_p(E_p)\left[\Im\left\langle\,\psi_n^{prior}|W_{An}\,|\psi_n^{prior}\right\rangle \right.\\ 
  &  +2\Re\left\langle\,\psi_n^{{HM}}|W_{An}|\psi_n^{prior}\right\rangle +\left.\left\langle\,\psi_n^{{HM}}|W_{An}\,|\psi_n^{{HM}}\right\rangle\right].
   \end{align}
The first term corresponds to elastic breakup followed by capture, while the third one contains all other processes involving the $n+A$ system. As we will show all three terms are, in general, important, and have to be simultaneously taken into account. For completeness, we present in Appendix B the partial wave decomposition of these results.

We have just shown that the non-orthogonality term in the  above expression
arises when disentangling  the elastic and 
non-elastic breakup contributions from $V_{prior}$. On the other hand, an identical non-orthogonality function
appears for different reasons in the derivation of the standard DWBA equations in the post form, but has been dropped because
of the equivalence of the final (proton channel) distorted waves $X_f$ and $\chi_f$ in the asymptotic region (see eq. (\ref{eq146})).
Nonetheless, in Eq.(\ref{eq34}) the proton distorted wave is certainly needed for small values of the proton coordinate, so the question may arise 
whether the latter non-orthogonality function should be kept after all  for its derivation. 

To see that such a term is actually not needed, let us check the consequences of  including it. By examining eq. (\ref{eq136}), it can be seen that this is equivalent to making the replacement:
\begin{align}\label{eq144}
V\rightarrow V_{pn}+\left(K_p+U_{Ap}+h_B-E\right)
          \end{align}
in  eq. (\ref{eq25}). We obtain from the post-form matrix element in Eq.(\ref{eq25}):
\begin{align}
\nonumber &\left\langle\,\phi_d\phi_A\chi_i|V_{pn}+\left(K_p+U^\dagger_{Ap}+h_B-E\right)\,|\chi_f^{(-)}\right)G_B\\
\nonumber &\times\left(\chi_f^{(-)}|(V_{pn}+\left(K_p+U_{Ap}+h_B-E\right))\,|\,\phi_d\phi_A\chi_i\right\rangle=\\
\nonumber&\bra{\phi_A}\left[S^*_{post}+\psi_n^{{HM}*}\left(h_B+E_p-E\right)\right]G_B\\
\nonumber&\times\left[S_{post}+\psi_n^{{HM}}\left(h_B+E_p-E\right)\right]\ket{\phi_A},\end{align}
where we have used eqs. (\ref{eq140}) and (\ref{eq35}). Taking into account that $\left(h_B+E_p-E\right)G_B=\idop$, we get
\begin{align}\label{eq156}
\nonumber S^*_{post}&G_B^{opt}S_{post}+2\Re\left(S_{post}\psi_n^{{HM}}\right)\\&+\bra{\phi_A}S^*_{post}\left(h_B+E_p-E\right)S_{post}\ket{\phi_A}.
\end{align}
When taking the imaginary part, only the first term of the above expression survives, and we obtain again Eq.(\ref{eq32}).
We conclude that the post source term defined in eq. (\ref{eq141}) should not include a non--orthogonality contribution.

%%%%%%%%%%%%%%%%%%%%%%%%%%%%%%%%%%%%%%%%%%%%%%%%%%%%%%%%%%%%%%%%%%%%%%%%%%%%%%%%%%%%%%%%%%%%%%%%%

\subsection{Transfer to bound states}
\label{theory-bound}

The sort of breakup processes we are considering can be thought of as transfer to the continuum. One would like to have one  framework to describe both transfer to bound states and continuum states. The Green's function formalism allows for this connection. This idea was first proposed in \cite{udagawa1987}.
We write the partial wave coefficients of the Green's function $G_B^{opt}(\mathbf r_{An},\mathbf r_{An}')$ defined in Eq.(\ref{eq31}), as 
 \begin{equation}\label{eq65}
G_l(r_{An},r_{An}')= \frac{f_l(k_n,r_{An<})g_l(k_n,r_{An>})}{k_n r_{An}r_{An}'},
\end{equation}
where $k_n=\sqrt{2 m_n \varepsilon}/\hbar$, and $f_l(k_n,r_{An})\, (g_l(k_n,r_{An}))$ is the regular (irregular) solution of the homogeneous equation
\begin{align}\label{eq67}
\left(-\frac{\hbar^2}{2m_n}\frac{d^2}{dr_{An}^2}+U_{An}(r_{An})+\frac{\hbar^2 l(l{+}1)}{2m_n r_{An}^2}-\varepsilon\right)\,\{f_l ,g_l\}(k_n,r_{An})=0.
\end{align}
At the origin we impose $\lim_{r_{An}\rightarrow 0}f_l(k_n,r_{An})=0$ for the regular solution. At large distances the boundary condition of course depends on whether the energy $\varepsilon$ is positive or negative. For scattering neutron states (positive $\varepsilon$),
 \begin{align}\label{eq70p}
\lim_{r_{An}\rightarrow \infty}g_l(k_n,r_{An})\rightarrow e^{i(k_nr_{An}-\frac{l\pi}{2})}, 
 \end{align}
 while for final neutron bound states (negative $\varepsilon$),
  \begin{align}\label{eq70n}
 \lim_{r_{An}\rightarrow \infty}g_l(k_n,r_{n})\rightarrow e^{-(\kappa_nr_{An})}, 
  \end{align}
with  $\kappa_n=\sqrt{-2m_n\varepsilon}/\hbar$.

%%%%%%%%%%%%%%%%%%%%%%%%%%%%%%%%%%%%%%%%%%%%%%%%%%%%%%%%%%%%%%%%%%%%%%%%%%%%%%%%%%%%%%%%%%%%%%%%%

%\subsection{Negative energies and direct one--particle transfer}
If the imaginary part $W_{An}$ of the neutron--target optical potential  is small, we can use first order perturbation theory to express
\begin{align}\label{eq114}
G_B^{opt}(\mathbf r_{An},\mathbf r_{An}';E)\approx\frac{\hbar^2}{2m_n}\sum_n\frac{\phi^*_n(\mathbf r_{An}')\phi_n(\mathbf r_{An})\,}{E_n+i\Gamma_n/2-E},
\end{align}
where
\begin{align}\label{eq115}
\Gamma_n=2\int \phi^*_nW_{An}\phi_n\,d\mathbf r_{An},
\end{align}
and $\phi_n, E_n$ are the eigenfunctions and eigenvalues of the Schr\"odinger equation corresponding to the \emph{real} part of the optical potential. If we now consider an energy $E$ close to an isolated resonance, i.e. $|E-E_n|\ll|E_m-E_n|$, only the $n$th term of the sum will contribute to the Green's function, and
\begin{align}\label{eq116}
G_B^{opt}(\mathbf r_{An},\mathbf r'_{An};k)\approx\frac{\hbar^2}{2m_n}\frac{\phi^*_n(\mathbf r'_{An})\phi_n(\mathbf r_{An})\,}{E_n+i\Gamma_n/2-E}.
\end{align}
The resulting neutron wave function is
\begin{align}\label{eq117}
\psi(\mathbf r_{An})= \frac{\hbar^2}{2m_n}\frac{\phi_n(\mathbf r_{An})}{E_n+i\Gamma_n/2-E}\int \phi^*_n(\mathbf r'_{An})
   S(\mathbf r'_{An})d\mathbf r'_{An}.
\end{align}
According to the particular nature of the source term (see Eqs. (\ref{eq141}) and (\ref{eq57})), the integral in Eq. (\ref{eq117}) has the form of a one--neutron transfer DWBA amplitude 
\begin{align}\label{eq150}
T^{\rm (1NT)}_n=\int \phi^*_n\, S_n \, d\mathbf r'_{An}=\int \phi^*_n\left(\,\chi_f^{(-)}\big |V_{post,prior}\big |\phi_d\,\chi_i\right\rangle d\mathbf r'_{An}
\end{align}
to the single--particle state $\phi_n$ of the target--neutron residual nucleus. We can then write:
\begin{align}\label{eq118}
\psi_n(\mathbf r_{An})= \frac{T^{\rm (1NT)}_n}{E_n+i\Gamma_n/2-E}\phi_n(\mathbf r_{An}),
\end{align}
and the final neutron wavefunction $\psi_n(\mathbf r_{An})$  can be interpreted as the $n$th eigenstate of the neutron--target (real) single--particle potential times the direct transfer amplitude to this particular state, modulated by an energy denominator. 
The absorption cross section is proportional to the matrix element
\begin{align}\label{eq119}
\nonumber \langle \psi|W_{An}|\psi\rangle= &\frac{\left|T^{\rm (1NT)}_n\right|^2}{\left(E_n-E\right)^2+\Gamma_n^2/4}\int \phi^*_nW_{An}\phi_n\,d\mathbf r_{An}\\
&=\frac{1}{2}\frac{\Gamma_n}{\left(E_n-E\right)^2+\Gamma_n^2/4}\left|T^{\rm (1NT)}_n\right|^2.
\end{align}
As a consequence, if the transfer amplitude $T^{\rm (1NT)}_n$ is approximately constant in an energy interval of the order of $\Gamma_n$, the  energy--dependent differential cross section around a resonance has a Lorentzian shape, and the integrated cross section under the peak  is independent of $W_{An}$ for small enough $W_{An}$.

 It can be shown (see Appendix C) that  there is a simple relationship between the cross section for the capture of a neutron in a bound state of finite width and the cross section for direct transfer to the corresponding zero--width bound state. Assuming again that $T^{\rm (1NT)}_n$ is essentially constant in an energy range of the order of $\Gamma_n=2\langle W_{An}\rangle$, we have
 \begin{align}\label{eq124}
 \left.\frac{d^2\sigma}{d\Omega_pdE_p}(E,\Omega)\right]^{NEB}\approx\frac{1}{2\pi}\frac{\Gamma_n}{\left(E_n-E\right)^2+\Gamma_n^2/4}\frac{d\sigma_n}{d\Omega}(\Omega),
 \end{align}
where $\frac{d\sigma_n}{d\Omega}$ is the direct transfer differential cross section to the $n$th eigenstate of the real potential. At the resonance energy  peak $(E=E_n)$, we have the simple relationship
 \begin{align}\label{eq125}
 \left.\frac{d^2\sigma}{d\Omega_pdE_p}(E=E_n,\Omega)\right]^{NEB}\approx\frac{2}{\Gamma_n\pi}\frac{d\sigma_n}{d\Omega}(\Omega).
 \end{align}

%%%%%%%%%%%%%%%%%%%%%%%%%%%%%%%%%%%%%%%%%%%%%%%%%%%%%%%%%%%%%%%%%%%%%%%%%%%%%%%%%%%%%%%%%%%%%%%%%
\section{Results}
\label{results}

\subsection{Numerical details}

\begin{table}
\begin{center}
\begin{tabular}{|c|c|c|c|c|c|c|c|c|}
\hline  & $V$ & $W$ & $W_D$ & $a$ & $a_D$ & $r	$ & $r_D$ & $r_C$ \\ 
\hline $d$ & 99.0 & 0.0 & 16.7 & 0.84 &  0.64&  1.12&  1.31&  1.30\\ 
\hline $p$ & 50.6 & 0.0 & 14.1 & 0.678 &  0.47&  1.25&  1.25&  1.25\\
\hline
\end{tabular}
\caption{Optical model parameters. Energies are expressed in MeV and lengths in fm. In the proton channel, the parameters are those listed in \cite{perey76} for proton--$^{93}$Nb scattering at 16.2 MeV, while in the deuteron channel they correspond to those listed in the same reference for deuteron--$^{93}$Nb scattering at 17 MeV.}\label{tab1}
\end{center}
\end{table}

We present the results obtained for the reaction $^{93}$Nb($d,p$) at two different beam energies, $E_d=15$ MeV, and $E_d=25.5$ MeV. The optical model potentials $U_{Ad}$, $U_{Ap}$ used in the initial (deuteron) and final (proton) channels respectively were taken from the Perey and Perey compilation (\cite{perey76}), and are summarized in Table \ref{tab1}. For the final state interaction $U_{An}$ between the neutron and the $^{93}$Nb target, we have slightly modified the energy dependent Koning--Delaroche global optical nucleon--nucleus potential (\cite{koning03}), by setting the real part parameter to $V=50.3$ MeV in the whole energy range, and keeping the original energy dependence of the other parameters. This was needed to reproduce essential features of the $^{94}$Nb nucleus. Furthermore, although we respect the energy dependence of the depth of the surface imaginary part $W_D$ of the Koning-Delaroche potential,  we do not let it fall below 4 MeV, corresponding to the experimental energy resolution of the data in \cite{mastroleo1990}.
 The maximum partial wave $l_p$ used in the calculations is 15 and 20 for $E_d=15$ MeV and $E_d=25.5$ MeV respectively. The contribution of final neutron states with $l\geq 8$ is found to be very small.     
 
The  deuteron ground state wavefunction is taken to be an $L=0$ state with a radial wavefunction generated by a Woods--Saxon potential with radius $R_d=0.4$ fm and diffusivity $a_d=0.6$ fm. When the real depth is adjusted to reproduce the binding energy of deuteron, the resulting wavefunction is compatible with the experimental value of the mean square radius of the deuteron and the zero-range constant $D_0 = -122.5 $ MeV.fm$^{3/2}$.  

\begin{figure}[t!]
\begin{center}
\includegraphics[width=0.90\columnwidth]{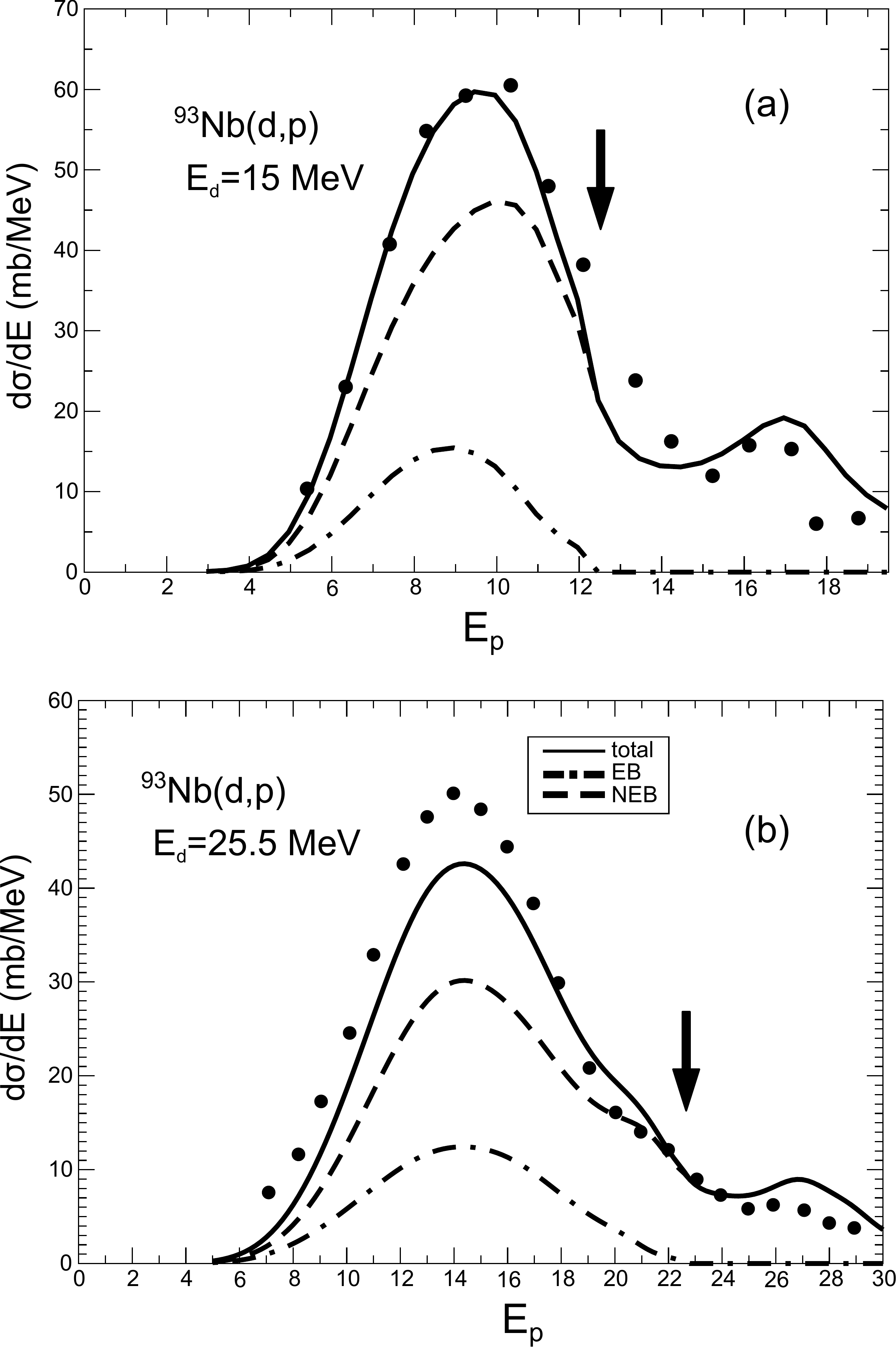}
\end{center}
\caption{Energy distributions of the detected proton for $^{93}$Nb(d,p) at (a) $15$ MeV and (b) $25.5$ MeV: total cross sections (solid line), elastic breakup (EB, dot-dashed line) and inelastic breakup (NEB, dashed line). The arrows indicate the position of the neutron emission threshold. Data is from \cite{mastroleo1990}.}
\label{fig:energy}
\end{figure} 

In Fig.\ \ref{fig:energy} we show the proton energy distributions for $^{93}$Nb(d,p) at $15$ MeV and $25.5$ MeV along with the data \cite{mastroleo1990}. Also shown is the breakdown into elastic breakup (EB) and inelastic breakup (NEB). Our results indicate that the inelastic breakup is dominant at all energies, nevertheless elastic breakup is important particularly  around the peak of the distribution. The comparison with the data demonstrates that our model provides a good account of the process. A close comparison with the theoretical results presented in \cite{mastroleo1990} show significant differences that can be partially attributed to the neglect of the additional terms arising from non-orthogonality. 

We also compared our results with those from \cite{lei2015}.  At the peak of  our distribution for the higher beam energy ($E_p=14$ MeV), the  elastic breakup contributes  $25$\% of the total cross section. This differs significantly from the results presented in \cite{lei2015}, where the elastic breakup is less that $10$\% of the total cross section at the peak. This is an important issue because elastic breakup will not lead to the neutron being captured into a compound state and therefore will need to be subtracted from the total cross section in order to apply the surrogate method. In the work of \cite{lei2015}, the elastic breakup is treated in a separate formalism, namely with the continuum discretized coupled channel method. In addition, given that the inelastic contribution is computed in the post formalism, the authors used continuum bins to address the convergence issues. A more detailed comparison between these two methods will be very useful. 

\begin{figure}[t!]
\begin{center}
\includegraphics[width=0.90\columnwidth]{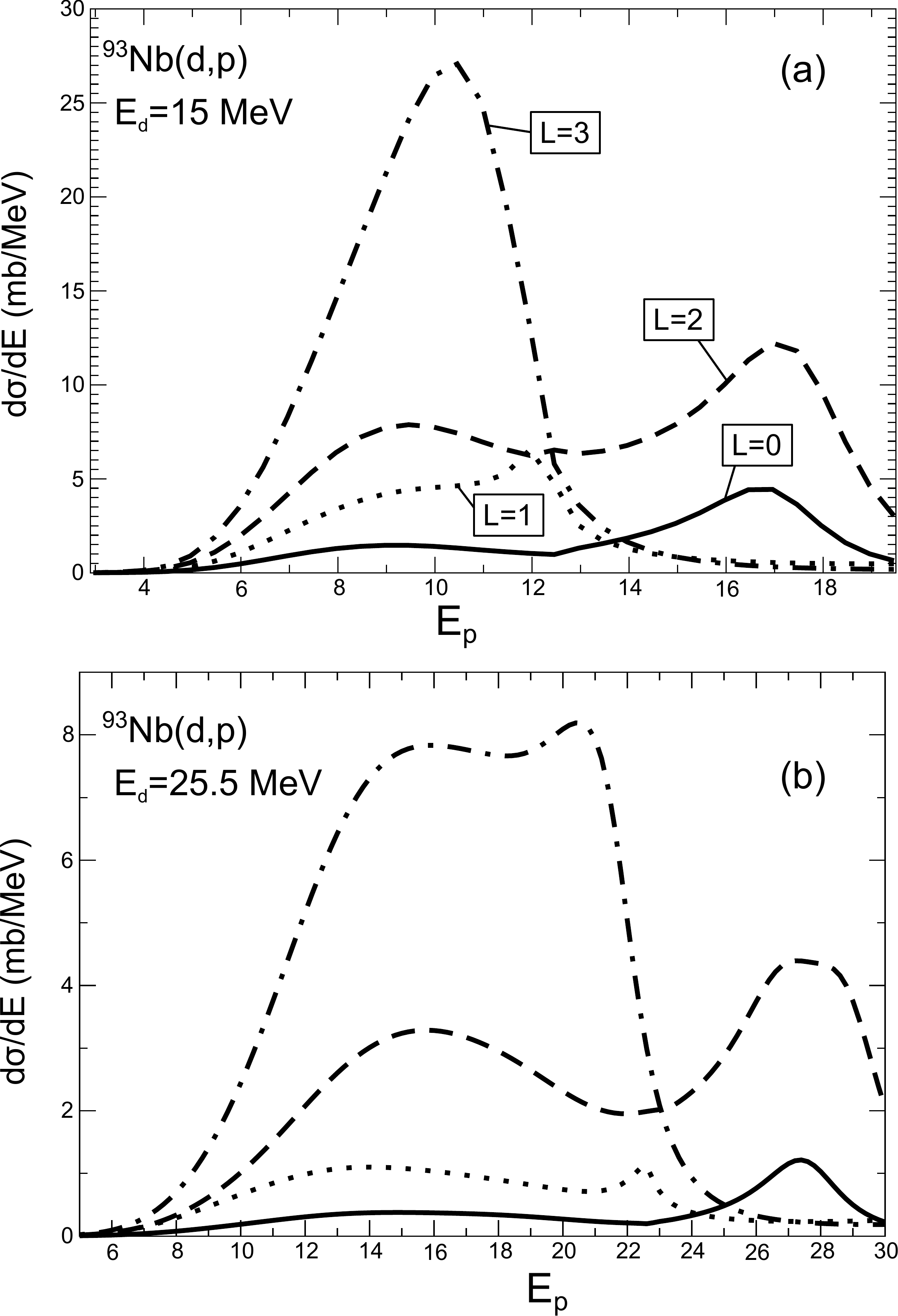}
\end{center}
\caption{Energy distributions for $^{93}$Nb(d,p) at (a) $15$ MeV and (b) $25.5$ MeV: partial wave decomposition .}
\label{fig:partial}
\end{figure} 

For the surrogate method,  the spin distributions of the (d,p) cross sections are important in making the connection to neutron capture (see Section \ref{intro}).
In Fig. \ref{fig:partial} we provide the breakdown in terms of the various angular momenta, for both beam energies considered.  A strong peak is found around $E_p=10$ MeV, for $L=3$ corresponding to a narrow resonance in the neutron-target system. A much broader resonance is present in the $L=1$ channel and therefore the peak in that component is less pronounced. For $L=0$ and $L=2$, we can see the signature of the neutron-target bound states, at high proton energy. From the distributions in Fig.\ \ref{fig:partial}, it is obvious that the Weisskopf approximation, typically used in the analysis of surrogate reactions, is not valid. The cross section is strongly dependent on the angular momentum and closely connected to the internal structure of the composite final state.

\begin{figure}[t!]
\begin{center}
\includegraphics[width=0.90\columnwidth]{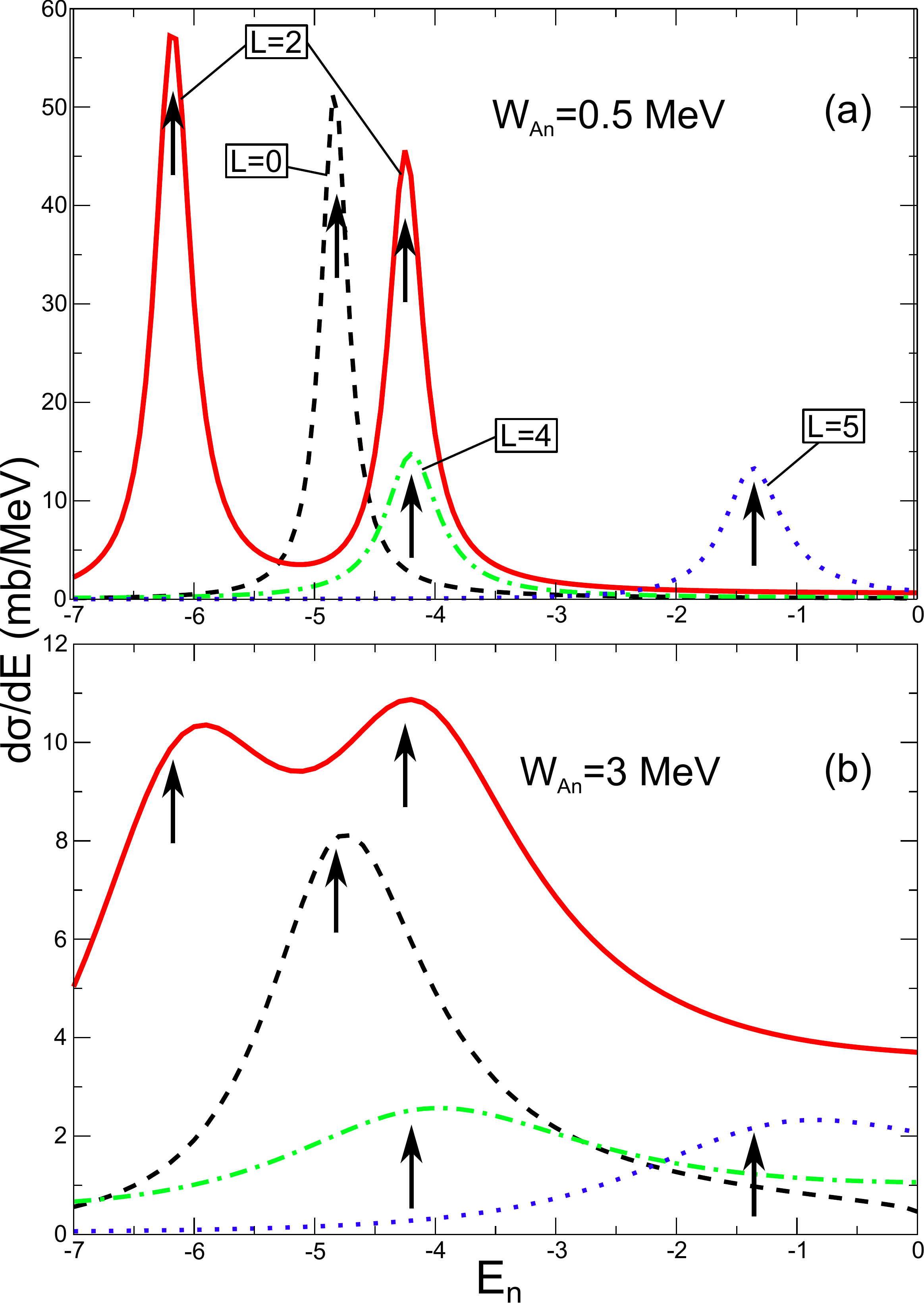}
\end{center}
\caption{Energy distributions for $^{93}$Nb(d,p) at  $15$ MeV below the neutron threshold, for those partial waves that have bound states: (a)  strength of the imaginary potential $W=0.5 MeV$. and (b) strength of the imaginary potential $W=3.0$ MeV.}
\label{fig:bound}
\end{figure} 
One attractive feature of the model here developed is the ability to provide, in a consistent framework, a description of neutron capture into the continuum and into bound states. The optical potential used to describe the n-target system is primarily based on fits to elastic scattering, appropriate for positive energies, but our results depend on the continuation of the potential  to negative energies. In Figs.\ \ref{fig:energy} and \ref{fig:partial}, for the highest proton energies, corresponding to the neutron below threshold, we keep $W_D=4$ MeV, as stated above, to describe limited experimental resolution. In Fig.\ \ref{fig:bound}, by contrast, we explore the effects of smaller imaginary terms in $U_{An}$ in the $^{93}$Nb(d,p) reaction, by changing this value to $W_D=0.5$ MeV (a) and $W_D=3$ MeV (b). For the smallest imaginary term, bound states appear as narrow peaks. Increasing the imaginary term increases their widths. An example of this can be seen for the $L=2$ states: when the imaginary term is small, we can resolve the $L=2$ spin-orbit partners, while as we increase the imaginary term, these two states become blurred to the point of becoming indistinguishable in Fig.\ \ref{fig:partial}.  The arrows in Fig.\ \ref{fig:partial} correspond to the location of the bound states for the real potential. We find that that in addition to the spreading of the states, the introduction of the imaginary component can also shift the peaks to higher energy. As discussed in section \ref{theory-bound}, the limit of $W_D=0$ MeV in our formulation corresponds to the standard DWBA transfer to bound states. We have indeed used this fact to check our calculations by comparing our results in this limit to those produced by {\sc Fresco} \cite{fresco}. Complete agreement was found.

\begin{figure}[t!]
\begin{center}
\includegraphics[width=0.90\columnwidth]{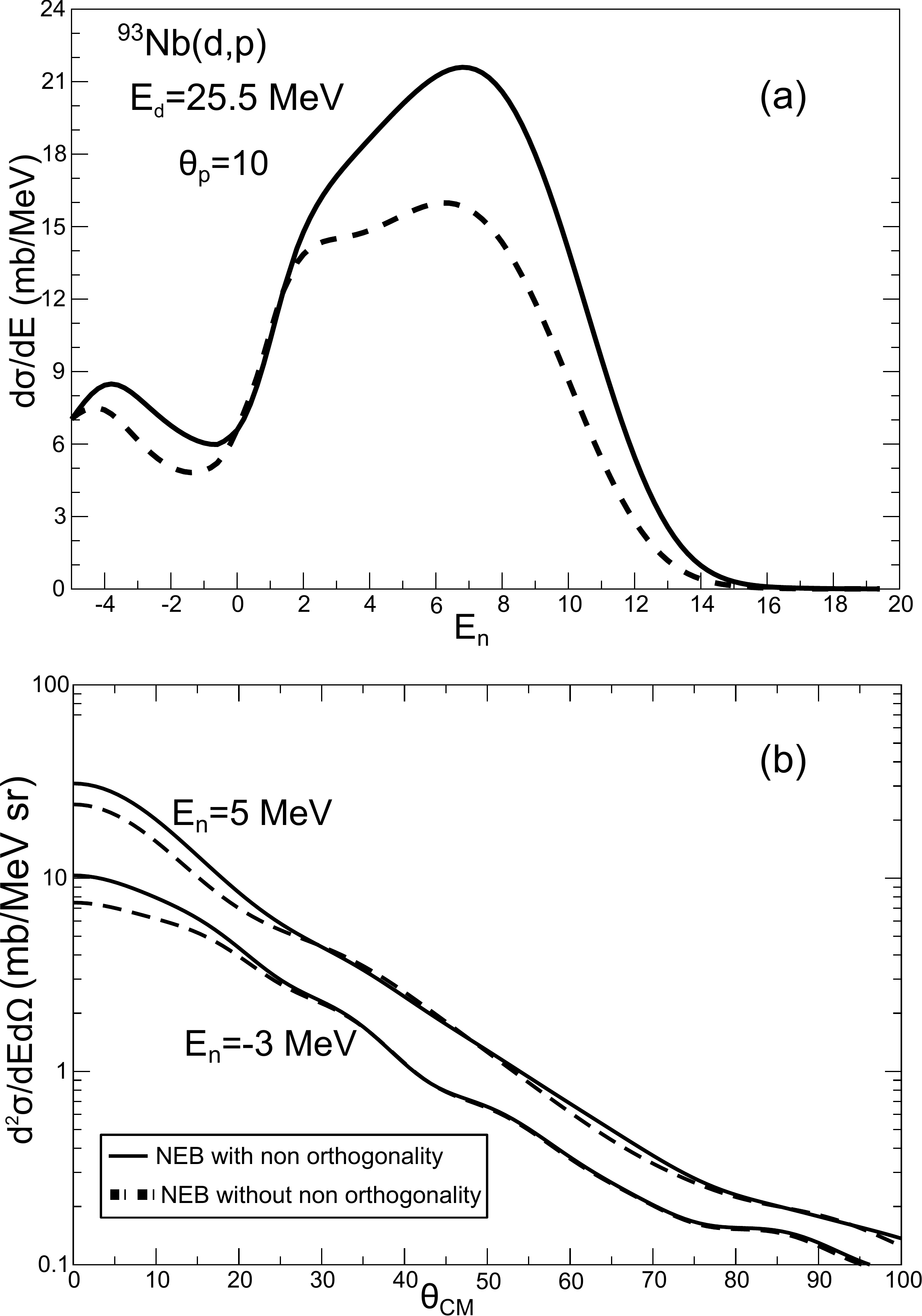}
\end{center}
\caption{Effect of the non-orthogonality and cross terms in the cross sections: (a) energy distribution for $^{93}$Nb(d,p) at  $25.5$ MeV  for an angle $\theta=10 \circ$ and (b) angular distributions for both $E_n=-3$ MeV and $E_n=5$ MeV. }
\label{fig:no}
\end{figure} 
The basis of the large controversy between the AV and the UT approaches stems from the non-orthogonality term, discussed in detail in section \ref{theory-UT}. It is therefore critical to understand the importance of this term in the calculations. In Fig.\ \ref{fig:no} we plot the inelastic breakup cross sections, with (solid line) and without  (dashed line) the non-orthogonality term. The neutron energy distribution, for $\theta_p =10^\circ$, is shown in panel (a) and the angular distribution, for two neutron energies, is shown in panel (b). The non-orthogonality term is most important at the peak of the energy distribution but remains important even when the neutron is captured into bound states. In terms of the angular distribution, the non-orthogonality term manifests itself mostly at forward angles. The results in Fig.\ \ref{fig:no} demonstrate the need for the inclusion of the non-orthogonality term in the prior-form formalism. The discrepancies found in the literature between the AV and the UT approaches are, to a large extent, a consequence of the fact that UT neglects the non-orthogonality term.

%%%%%%%%%%%%%%%%%%%%%%%%%%%%%%%%%%%%%%%%%%%%%%%%%%%%%%%%%%%%%%%%%%%%%%%%%%%%%%%%%%%%%%%%%%%%%%%%%%%%%%%%%%%%%%%%%%%%%%%%%%%%%%%%%%%

\section{Conclusions}
\label{conclusions}

We have derived, implemented and validated a practical theory for computing cross sections for inclusive processes of the type A(d,p)X,  where only the proton is detected in the final state. This includes elastic breakup as well as all other inelastic processes. Our model is based on the assumption that the proton is a spectator and cannot  excite the target. The same framework is able to describe these reactions across the full neutron energy range in a consistent manner. Our model is directly relevant to the application of the surrogate method for extracting neutron capture cross sections from (d,p) reactions. The present formalism can be generalized to describe the partial fusion of any cluster in a loosely bound projectile. 

We discuss in detail the post and the prior formalism. While the cross sections derived in the post-form are formally more elegant, they pose convergence challenges. This challenge is removed in the prior formalism. However, additional terms due to non-orthogonality arise. We show that these terms are important and cannot be neglected for all energies of the detected proton.  We also predict the spin distribution of the resulting compound nucleus, an important ingredient for the application of the surrogate method to extract neutron capture reactions. We show that the cross sections are strongly dependent on the angular momentum, a factor that needs to be incorporated in the analysis of the surrogate data. Finally, we also make predictions regarding the relative contribution of the elastic breakup and all other inelastic processes. The elastic breakup process does not lead to neutron capture and needs to be subtracted from the total (d,p) cross section when relating it back to (n,$\gamma$).
In this work, we have resolved a decades-old controversy, namely that between Austern {\it et al.} and Udagawa {\it et al.}, promulgating respectively the post and the prior methods for describing the inclusive process $A(d,p)X$. 

While our method is able to describe the one example we have studied, it is important to perform a systematic study for various nuclei across the nuclear chart. Also, since our method makes consistent predictions for both the elastic and inelastic breakup, it would be very useful to have the measurement of elastic breakup for the same cases that have been studied with the inclusive experiments.

In comparing our methods with other calculations, we found small differences with \cite{mastroleo1990}, which can be partially attributed to the neglect of the non-orthogonality induced terms. More concerning, we found significant differences with the predictions of \cite{lei2015}, particularly in the elastic breakup cross section. Given that in \cite{lei2015} the elastic breakup was obtained through a different formalism, the comparison is not straightforward. Nevertheless, a better understand on the source of the differences is desirable. Again, elastic breakup data for these cases would be useful.

Obtaining a reliable  cross section for processes of the type A(d,p)X is only the first step in the application of the surrogate method. The elastic breakup cross section needs to be subtracted from the total cross section, and the remaining cross section needs to be treated within a statistical approach to determine the fraction of neutrons that end up being captured versus those that evaporate. The coupling to a statistical method is planned for the near future.

%%%%%%%%%%%%%%%%%%%%%%%%%%%%%%%%%%%%%%%%%%%%%%%%%%%%%%%%%%%%%%%%%%%%%%%%%%%%%%%%%%%%%%%%%%%%%%%%%

\begin{center}
\textbf{ACKNOWLEDGEMENT}
\end{center}

We are grateful to Jutta Escher for useful discussions. This work was supported by the National Science
Foundation under Grant No. PHY-1403906,   and the Department
of Energy, Office of Science, Office of Nuclear Physics under award No. DE-FG52-08NA28552,
and by Lawrence Livermore National Laboratory under Contract DE-AC52-07NA27344.

\newpage

\section*{Appendix A: Discussion of EB and NEB in the context of flux conversation}

We rewrite the differential equation for the neutron wavefunction eq. (\ref{eq35}) in the following form:
  \begin{align}\label{eq151}
\left(\nabla^2+\bar U_{An}-\bar E\right)\psi^{post}_{n}=\bar S_{post},
  \end{align}
 where 
    \begin{align}\label{eq155}
 \bar S_{post}=\frac{2m_n}{\hbar^2}S_{post},
    \end{align}
and $\bar U_{An}=\tfrac{2m_n}{\hbar^2}U_{An}$, $\bar E=\tfrac{2m_n}{\hbar^2}\left(E-E_p-\varepsilon_A\right)$.
From (\ref{eq151}) and its complex conjugate, and multiplying respectively by $\psi^{post*}_{n}$ and $\psi^{post}_{n}$ we get (we will drop the $post$ suffix)
   \begin{align}\label{eq152}
\nonumber \psi^{*}_{n}\left(\nabla^2+\bar U_{An}-\bar E\right)\psi_{n}&=\psi^{*}_{n} \bar S\\
 \psi_{n}\left(\nabla^2+\bar U^\dagger_{An}-\bar E\right)\psi^{*}_{n}&=\psi_{n} \bar S^\dagger.
   \end{align}
 Substracting the above equations, and integrating over a large volume, we obtain
    \begin{align}\label{eq153}
\int\nabla\left(\psi_{n}\nabla\psi^{*}_{n}-\psi^{*}_{n}\nabla\psi_{n}\right)d\mathbf{r}_{n}-2i\int|\psi_{n}|^2 \bar W_{An}d\mathbf{r}_{n}= \nonumber \\
 = 2i\int \Im(\psi_{n} S^\dagger)d\mathbf{r}_{n}.
    \end{align}
    The first term can be cast into an outgoing elastic flux across the surface enclosing the volume, while the second term accounts for the non--elastic breakup. The above equation describes how the flux generated by the right--hand side term is converted into an elastic and a non--elastic contribution.

\section*{Appendix B: Partial wave decomposition of the differential cross section}
After partial wave decomposition, the multiple differential cross section can thus be written  in the prior and post forms,
          \begin{align}
          \label{eq83}
         \frac{d^2\sigma}{d\Omega_pdE_p}=\frac{2\pi}{\hbar v_d}\rho_p(E_p) \sum_{l,m} B_{lm}^ {\rm prior,post}
         \end{align}
  where the post contributions are        
        \begin{align}
          B^{\rm post}_{lm} = \int \big |\sum_{l_p}\phi^{post}_{lml_p}(r_{Bn};k_{p})Y^{l_p}_{-m}(\theta_p)\big|^2W_{An}(r_{An})\, \,dr_{Bn},
          \end{align}
     and the prior contributions
             \begin{multline}
          \label{eq85}
  B^{\rm prior}_{lm}=   
            \int \big |\sum_{l_p}\phi^{prior}_{lml_p}(r_{Bn};k_{p})Y^{l_p}_{-m}(\theta_p) \big |^2W_{An}(r_{An})\, \,dr_{Bn}\\
          +\int \big |\sum_{l_p}\phi^{{HM}}_{lml_p}(r_{Bn};k_{p})Y^{l_p}_{-m}(\theta_p) \big |^2W_{An}(r_{An})\, \,dr_{Bn}\\
           -2\Re \int \sum_{l_p,l_p'} \phi^{{HM}}_{lml_p}(r_{Bn};k_{p})\phi^{prior*}_{lml_p'}(r_{Bn};k_{p})\\
           \times Y^{l_p}_{-m}(\theta_p)Y^{l_p'}_{-m}(\theta_p)W_{An}(r_{An})\, \,dr_{Bn}.
          \end{multline}
Here we have used the neutron partial wave  functions, equivalently for prior or post,
 \begin{equation}
\psi_n(\mathbf{r}_{Bn};\mathbf{k}_{p})= \sum_{l,m,l_p}\phi_{lml_p}(r_{Bn};{k}_{p})Y^l_m(\theta_{Bn})Y^{l_p}_{-m}(\theta_p)/r_{Bn},
\end{equation} 
with the  source terms having a similar decomposition
\begin{align}\label{eq15}
S(\mathbf{r}_{n};\mathbf{k}_p)=\sum_{lml_p} s_{lml_p}(r_{n};\mathbf{k}_{p})Y_{lm}(\theta_{Bn})Y_{l_pm}(\theta),
\end{align}
so
\begin{align}\label{eq68}
\nonumber \phi_{lml_p}(r_{An},k_p)=&\int G_l(r_{An},r_{An}') \, 
  s_{lml_p}(r_{An}';k_p)\, r_{An}'^2dr_{An}'\\
\nonumber=&\frac{1}{k_n}\left(g_l(k_n,r_{An})\int_0^{r_{An}}f_l(k_n,r_{An}')s_{lml_p}(r_{An}';k_p)\, r_{An}'dr_{n}'\right.\\
&\left.+f_l(k_n,r_{An})\int_{r_{An}}^{\infty}g_l(k_n,r_{An}') s_{lml_p}(r_{An}';k_p)\, r_{An}'dr_{An}'\right).
\end{align}

\section*{Appendix C: Transfer cross sections to bound states}

In this appendix we show that the value of the integrated cross section under the peak in Eq.(49) is equal to the direct transfer cross section to a sharp bound state. Let's  consider  the limit 
\begin{align}\label{eq120}
\lim_{W_{An}\rightarrow 0}\langle \psi|W_{An}|\psi\rangle=\pi\left|T^{\rm (1NT)}_n\right|^2\delta(E_n-E),
\end{align}
and substitute (\ref{eq120}) in Eq. (\ref{eq34}), 

 \begin{align}\label{eq121}
 \nonumber \left.\frac{d^2\sigma}{d\Omega_pdE_p}\right]^{NEB}&=-\frac{m_dm_p}{4\pi^3\hbar^4}\frac{k_p}{k_d}\left\langle\psi_{n}\right|  W_{An}\, \left|\psi_{n}\right\rangle\\
 &=\frac{m_dm_p}{4\pi^2\hbar^4}\frac{k_p}{k_d}\left|T^{\rm (1NT)}_n\right|^2\delta(E_n-E),
 \end{align}
 where the subindices $p,d$ refer to the proton and the deuteron respectively, and we have used the density of levels (\ref{eq82}).
If we integrate over a vanishing interval $\delta E$ around $E_n$, we get
 \begin{align}\label{eq122} \left.\frac{d\sigma}{d\Omega_p}\right]^{NEB}=\frac{m_dm_p}{4\pi^2\hbar^4}\frac{k_p}{k_d}\left|T^{\rm (1NT)}_n\right|^2,
 \end{align}
which is the DWBA transfer differential cross section. 

Moreover, in this limit the non--orthogonality term and the cross term  vanish. Indeed, the functions  $\varphi^{{HM}}_{lml_p}(r_{Bn};k_{p})$, not being affected by the propagator, do not exhibit any resonant behavior around the poles $E_n$. Thus the  second term in Eq. (\ref{eq85}) vanishes as $W_{An}\rightarrow 0$. On the other hand, the third term behaves as
 \begin{align}\label{eq123}
\sim\lim_{\Gamma_n\rightarrow 0}\frac{\Gamma_n}{(E_n-E)+i\Gamma_n/2}
 \end{align}
when $W_{An}\rightarrow 0$, and is equal to zero everywhere except in the zero--measure interval around $E_n=E$, where it has the finite value $-2i$. Its contribution thus vanishes after the energy integration. In addition, the interaction $U_{An}$ used in the prior representation (see Section \ref{theory-UT}) coincides now with the real potential $V_{An}$, so the standard post-prior symmetry is fully recovered.

%\bibliography{fusion}

\end{document}